# Audiogmenter: a MATLAB Toolbox for Audio Data Augmentation

Gianluca Maguolo[a*], Michelangelo Paci[b], Loris Nanni[a], Ludovico Bonan[a]
[a] DEI, University of Padua, viale Gradenigo 6, Padua, Italy.
[b] BioMediTech, Faculty of Medicine and Health Technology, Tampere University, Arvo Ylpön katu 34, D 219, FI-33520, Tampere, Finland

*Corresponding author, email: gianluca.maguolo@phd.unipd.it

## Abstract

Audio data augmentation is a key step in training deep neural networks for solving audio classification tasks. In this paper, we introduce Audiogmenter, a novel audio data augmentation library in MATLAB. We provide 15 different augmentation algorithms for raw audio data and 8 for spectrograms. We efficiently implemented several augmentation techniques whose usefulness has been extensively proved in the literature. To the best of our knowledge, this is the largest MATLAB audio data augmentation library freely available. We validate the efficiency of our algorithms evaluating them on the ESC-50 dataset. The toolbox and its documentation can be downloaded at https://github.com/LorisNanni/Audiogmenter.

## Keywords

Audio augmentation; Data augmentation; Audio classification; Spectrogram; Convolutional neural network.

## 1. Introduction

Deep neural networks achieved state of the art performances in many artificial intelligence fields, like image classification [1], object detection [2] and audio classification [3]. However, they usually need a very large amount of labelled data to obtain good results and these data might not be available due to high labelling costs or due to the scarcity of the samples. Data augmentation is a powerful tool to improve the performance of neural networks. It consists in modifying the original samples to create new ones, without changing their labels [4]. This leads to a much larger training set and, hence, to better results. Since data augmentation is a standard technique that is used in most papers, a user-friendly library containing efficient implementations of these algorithms would be very helpful to researchers.

In this paper we introduce Audiogmenter, a MATLAB toolbox for audio data augmentation. In the field of audio classification and speech recognition, to the best of our knowledge, this is the first library specifically designed for audio data augmentation. Audio data augmentation techniques fall into two different categories, depending on whether they are directly applied to the audio signal [5] or to a spectrogram generated from the audio signal [6]. We propose 15 algorithms to augment raw audio data and 8 methods to augment spectrogram data. We also provide the functions to map raw audios into spectrograms. The augmentation techniques range from very standard techniques, like pitch shift or time delay, to more recent and very effective tools like frequency masking. The library is available at https://github.com/LorisNanni/Audiogmenter. The main contribution of this paper is to share a set of powerful data augmentation tools for researchers in the field of audio-related artificial intelligence tasks.

The rest of the paper is organized as follows: Section 2 describes the specific problem background and our strategy for audio data augmentation; Section 3 details the implementation of the toolbox; Section 4 provides one illustrative example; Section 5 contains experimental results; in Section 6 conclusions are drawn.

## 2. Related Work

To the best of our knowledge, Audiogmenter is the first MATLAB library specifically designed for audio data augmentation. Such libraries exist in other languages like Python. A well-known Python audio library is Librosa [7]. The aim of Librosa was to create a set of tools to mine audio databases, but the result was an even more comprehensive library useful in all audio fields. Another Python library is Musical Data Augmentation (MUDA) [8], which is specifically designed for audio data augmentation and is not suitable for more general audio related tasks. MUDA only contains algorithms for pitch deformations, time stretching and signal perturbation, but does not contain algorithms like pass filters that would not be useful for generating music data.
Some audio augmentation toolboxes are also available in MATLAB. A famous library is the TSM toolbox. It contains the MATLAB implementations of many time-scale modification (TSM) algorithms [9,10]. TSM algorithms allow to modify the speed of an audio signal without changing its pitch. They provide many algorithms to do that because it is not trivial to do while maintaining the audio plausible and every algorithm addresses the problem in a different way. It is clear that this toolbox can be used only on those audio tasks that do not heavily depend on the speed of the sounds.

Recently, the 2019b version of MATLAB included a built-in audio data augmenter for training neural networks. It contains very basic functions which have the advantage of being computed on every mini-batch during training, hence they do not use a large quantity of memory. However, they can only be applied to the input layers of recurrent networks.

On first approximation, an audio sample can be represented as an $M$ by $N$ matrix, where $M$ is the number of samples acquired at a specific frame rate (e.g. 44100 Hz), and $N$ is the number of channels (e.g. one for mono and more for stereo samples). Classical methods for audio classification consisted in extracting acoustic features, e.g. Linear Prediction Cepstral Coefficient or Mel-Frequency Cepstral Coefficients, to build feature vectors used for training Support Vector Machines or Hidden Markov Models [11]. Nevertheless, with the diffusion of deep learning and the growing availability of powerful Graphic Processing Units (GPUs) the attention moved towards the visual representations of audio signals. They can be mapped into spectrograms, i.e. graphical representations of sounds as functions of time and frequency, and then classified using Convolutional Neural Networks (CNN) [12]. Unfortunately, several audio datasets (especially in the field of animal sound classification) are limited, e.g. CAT sound dataset (2965 samples in 10 classes) [13], BIRD sound dataset (2762 samples in 11 classes) [14], marine animal sound dataset (1700 samples in 32 classes) [15], etc. Neural networks are prone to overfitting, hence data augmentation can strongly improve their performance.

Among the techniques used in the literature to augment raw audio signals, pitch shift, noise addition, volume gain, time stretch, time shift and dynamic range compression are the most common. Moreover, the Audio Degradation Toolbox provides further techniques such as clipping, harmonic distortion, pass filters, MP3 compression and wow resampling [16]. Furthermore, Sprengel et al. [5] showed the efficacy of augmentation by summing two different audio signals from the same class into a new signal. For example, if two samples contain tweets from the same bird species, their sum will generate a third signal still belonging to the same tweet class. Not only the raw audio signals, but also their spectrograms can be augmented using standard techniques [6], e.g. time shift, pitch shift, noise addition, Vocal Track Length Normalization (VTLN) [17], Equalized Mixture Data Augmentation (EMDA) [18], Frequency Masking [19] and Thin-Plane-Spline Warping (TPSW) [20].

## 3. Background and strategy

Given an audio dataset X with M classes and variable number of samples per class $X = \{x_{1,1}, \dots x_{n_1,1}, x_{1,2}, \dots x_{n_2,2}, \dots, x_{1,M}, \dots x_{n_M,M}\}$ , where $x_{i,j}$ represents a generic audio sample *i* from the class *j*, we propose to augment $x_{i,j}$ with techniques working on raw audio signals and to augment the spectrogram $S(x_{i,j})$ produced by the same raw audio signals.

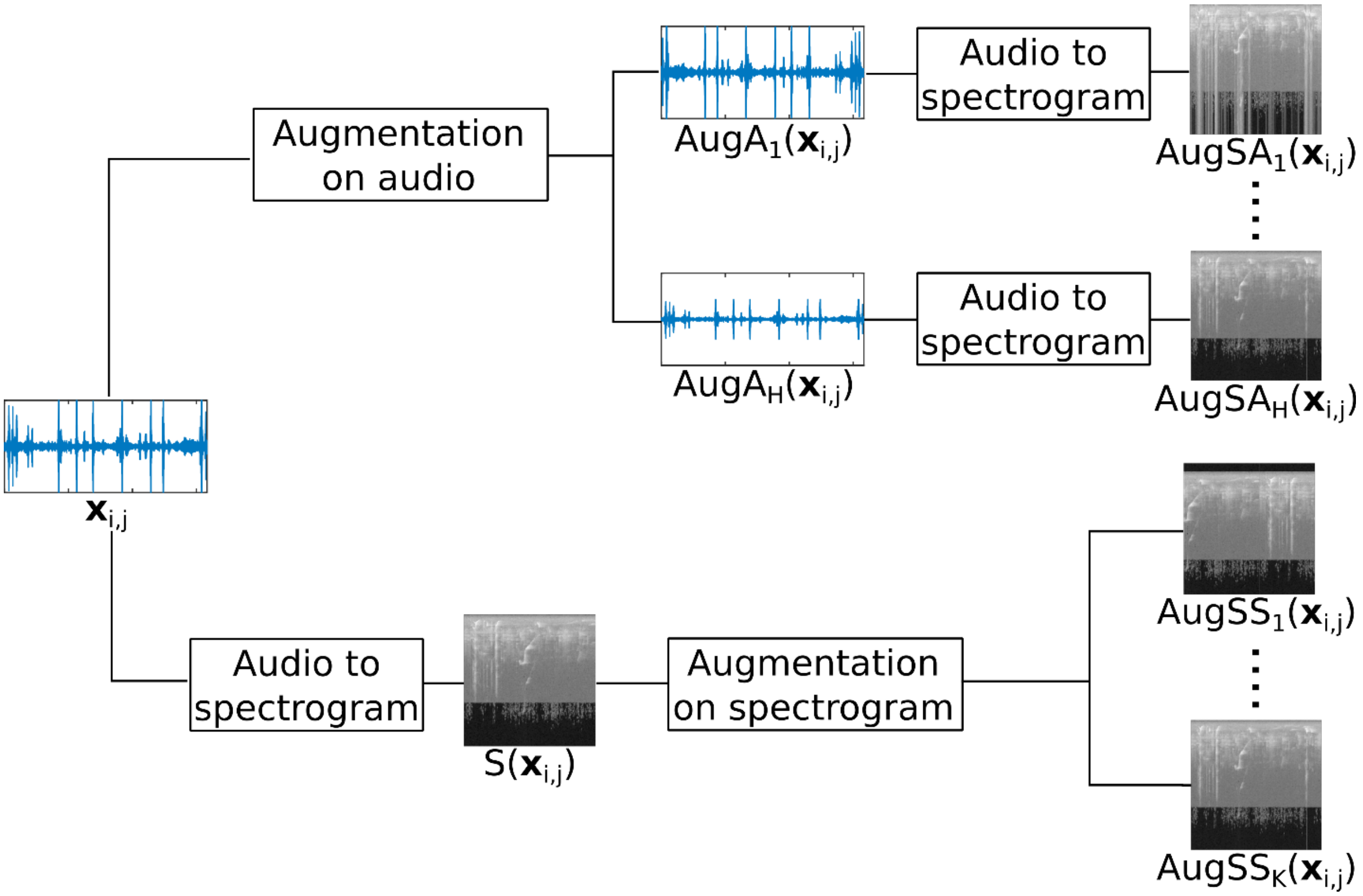

Figure 1. Augmentation strategy implemented in Audiogmenter. The upper branch shows how, from the original i-th audio sample $x_{i,j}$ from the class $j$, we obtain $H$ augmented audio samples $AugA_h(\overset{x}{\underset{¿}{¿¿}}i,j)$ to be converted into the augmented "Spectrograms from Audio" $AugSA_h(\overset{x}{\underset{¿}{¿¿}}i,j)$. The lower branch shows how $K$ augmented "Spectrograms from Spectrogram" $AugSS_k(\overset{x}{\underset{¿}{¿¿}}i,j)$ can be obtained from the spectrogram of the original audio sample $S(\overset{x}{\underset{¿}{¿¿}}i,j)$.

In our tool we used the function sgram included in the Large Time-Frequency Analysis Toolbox (LTFAT) [21], to convert raw audios into spectrograms.

Figure 1 depicts our strategy; from the original audio sample $x_{i,j}$ we obtain H intermediate augmented audio samples $AugA_h(\overset{x}{\underset{¿}{¿¿}}i,j)$ that are then converted into the "Spectrograms from Audio" $AugSA_h(\overset{x}{\underset{¿}{¿¿}}i,j)$; from the original spectrogram $S(\overset{x}{\underset{¿}{¿¿}}i,j)$ we obtain K augmented “Spectrograms from Spectrogram” $AugSS_k^x(i, j)$. The $H+K$ augmented spectrograms can then be used to train a CNN. In case of limited memory availability, one CNN can be trained with the $H$ $AugSA$ spectrograms, another with the $K$ $AugSS$ spectrograms and finally the scores can be combined by a fusion rule.

## 4. Toolbox structure and software Implementation

Audiogmenter is implemented as a MATLAB toolbox, using MATLAB 2019b. We also provide an online help as documentation (in the *./docs/* folder) that can be integrated into the MATLAB Help Browser just by adding the toolbox main folder to the MATLAB path.

The functions for the augmentation techniques working on raw audio samples are included in the folder *./tools/audio/*. In addition to our implementations of methods such as applyDynamicRangeCompressor.m and applyPitchShift.m, we also included four toolboxes, namely the Audio Degradation Toolbox by Mauch et al. [16], LTFAT [21], the Phase Vocoder from www.ee.columbia.edu/~dpwe/resources/matlab/pvoc/ and the Auditory Toolbox [22].

The functions for the augmentation methods working on spectrograms are grouped in the folder *./tools/images/*. In addition to our implementations of methods such as noiseS.m, spectrogramShift.m, spectrogramEMDA.m, etc., we included and exploited also a modified version of the code of TPSW [20].

Every augmentation method is contained in a different function. In *./tools/,* we also included the wrappers CreateDataAUGFromAudio.m and CreateDataAUGFromImage.m, using our augmentation techniques, respectively, from raw audio and spectrograms with standard parameters.

We now describe the augmentations and provide some suggestions on how to use them in the correct applications:

1. *applyWowResampling [16]* is similar to pitch shift but the intensity changes along time. The signal *x* is mapped into: $F(x) = x + a_m \frac{\sin(2\pi f_m x)}{2\pi f_m}$ where x is the input signal, and $a_m, f_m$ are parameters. This algorithm depends on the Degradation Toolbox. This is a very useful tool for many audio task and we recommend its use, although we suggest to avoid it for task that involve music, since changing the pitch with different intensity over time might lead to unnatural samples.
2. *addNoise* adds white noise to the input signal. It depends on the Degradation Toolbox. This algorithm improves the robustness of a tool by improving its performance on noisy signals, however this improvement might be unnoticed when the test set is not noisy. Besides, for tasks like sound generation one might want to avoid a neural network to learn from noise data.
3. *applyClipping* normalizes the audio signal leaving a percentage $X$ of the signal outside the interval [-1, 1]. Those parts of the signal are then mapped to sign(x). This algorithm depends on the Degradation Toolbox. Clipping is a common technique in audio processing, hence many recorded or generated audio might be played by a mobile device after having been clipped. If the tool the reader wants to train must recognize this kind of signal, we recommend this augmentation.

4. *applySpeedUp* modifies the speed of the signal by a given percentage. This algorithm depends on the Degradation Toolbox. We suggest to use this augmentation when the speed of a signal is not an important property of the signal.
5. *HarmonicDistortion* [16] applies the sine function to the signal multiple times. This algorithm depends on the Degradation Toolbox. This is a very specific augmentation that is not suitable for most applications. It is very useful to augment the input signals when the objective of the reader is working with sounds generated by electronic devices, since they might apply a small harmonic distortion to the original signal.
6. *applyGain* increases the gain of the input signal. We always recommend to use this algorithm, in general it can always be useful.
7. *applyRandTimeShift* randomly takes a signal $x(t)$ as input, where $0 \leq t \leq T$ . Then a random time $t^*$ is sampled and the new signal is $y(t) = x(mod(t+t^*, T))$ . In words, the first and the second part of the file are randomly switched. This algorithm is very useful, but do not use it if the order of the events in the input signals that you are working with is important. For example, it is not good for speech recognition. It is useful for tasks like sound classification.
8. *applySoundMix [23]* sums two audio signals from the same class. This algorithm depends on the Degradation Toolbox. We suggest to use this algorithm often. In particular, it is useful for multi-label classification or for tasks that involve multiple audio sources at the same time. It is worth noticing that is has also been used for single-label classification [24].
9. *applyDynamicRangeCompressor* applies, as its name says, Dynamic Range Compression [25]. This algorithm modifies the frequencies of the input signal. We refer to the original paper for a detailed description. Dynamic Range Compression is used to preprocess the audio before being played by an electronic device. Hence, a tool that deals with this kind of sounds should include this algorithms in its augmentation strategy.
10. *appltPitchShift* increase or decrease the frequencies of an audio file. This is one of the most common augmentation techniques. This algorithm depends on Phase Vocoder.
11. *applyAliasing* resamples the audio signal with a different frequency. It violates on purpose the Nyquist-Shannon sampling theorem [26] to degradate the audio signal. This is a modification of the sound that might occur when unsafely changing its frequency. This algorithm depends on the Degradation Toolbox. In general, it does not provide great improvement for machine learning tasks. We include it in our toolbox because it might be useful to reproduce the error due to the oversampling of low sampled signals, although they are quite rare in audio applications.
12. *applyDelay* adds a sequence of zeros at the beginning of the signal. This algorithm depends on the Degradation Toolbox. This time delay might be useful in any situation. In particular, we suggest to use it when the random shift of point 7 is not appropriate.
13. *applyLowpassFilter* attenuates the frequencies above a given threshold $f_1$ and blocks all the frequencies above a given threshold $f_2$ . This algorithm depends on the Degradation Toolbox. Low pass filters are useful when high frequencies are not relevant for the audio task.

14. *applyHighpassFilter* attenuates the frequencies below a given threshold $f_1$ and blocks all the frequencies below a given threshold $f_2$. This algorithm depends on the Degradation Toolbox. Low pass filters are useful when high frequencies are not relevant for the audio task.
15. *applyInpulseResponse [16]* modifies the audio signal as if it was produced by a particular source. For example, it simulates the distortion given by the sound system of a smartphone or it simulates the echo and the background noise of a great hall. This algorithm depends on the Degradation Toolbox. This augmentation is very useful if the reader needs to train a tool that must be robust and work in different environments.

The functions for spectrogram augmentation are:

1. *applySpectrogramRandomShifts* applies pitch and time shift. These augmentations are always useful.
2. *applySpectrogramSameClassSum [23]* sums the spectrograms of two images with the same label. This is a very useful algorithm. In particular, it is useful for multi-label classification or for tasks that involve multiple audio sources at the same time. It is worth noticing that is has also been used for single-label classification [24].
3. *applyVTLN* creates a new image by applying Vocal Tract Length Normalization (VTLN) [17]. For a more detailed description of the algorithm we refer to the original paper. Since vocal track length is one of the main inter-speaker differences in speech recognition, VTLN is particularly suited for this kind of applications.
4. *spectrogramEMDAaugmenter* applies Equalized Mixture Data Augmentation (EMDA) [18]. This function computes the weighted average of two randomly selected spectrograms belonging to the same class. It also applies a random time delay to one spectrogram and a perturbation to both spectrograms, according to the formula
$$s_{aug}(t)=\alpha\Phi(s_1(t),\psi_1)+(1-\alpha)\Phi(s_2(t-\beta T),\psi_2)$$
where $\alpha,\beta$ are two random numbers in $[0,1]$, $T$ is the maximum time shift and $\Phi$ is an equalizer function. We refer to the original paper for a more detailed description. This is a very general algorithm that works in very different situations. It is a more general version of *applySpectrogramSameClassSum.*
5. *applySpecRandTimeShift* does the same as *applyRandTimeShift*, but it works for spectrograms.
6. *randomImageWarp* applies Thin-Spline Image Warping [20] (TPS-Warp) to the spectrogram, on the horizontal axis. TPS-Warp consists in the linear interpolation of the points of the original image. In practice, it is a speed up where the change in speed is not constant and has average 1. This function is much slower than the others. It can be used in any application.
7. *applyFrequencyMasking* sets to a constant parameter the value of a some rows and some columns of the spectrogram. The effect is that it masks the real value of the input for randomly chosen times and frequencies. It was proposed in [3]. It was successfully used for speech recognition.

8. *applyNoiseS* adds noise to the spectrograms by multiplying the value of a given percentage of the pixels by a random number whose average is one and whose variance is a parameter. Similarly to *applyNoise,* this function increase the robustness of the trained tool on noisy data, however, if the test set is not noisy, the improvement might be unnoticed.

## 5. Illustrative Examples

In the folder *./examples/* we included testAugmentation.m that exploits the two wrappers detailed in the previous Section to augment six audio samples and their spectrograms, and plotTestAugmentation.m that shows the results from the previous function. The augmented spectrograms can be seen in Figures 2 and 3.
Figure 2 shows the effect of transforming the audio into a spectrogram and then applying the spectrogram augmentations described in the previous section. Figure 3 shows the spectrograms obtained by the original and the transformed audio files. Although these figures are different, it is possible to recognize specific features that are left unchanged by the augmentation algorithms, as it is desired for this kind of algorithms.

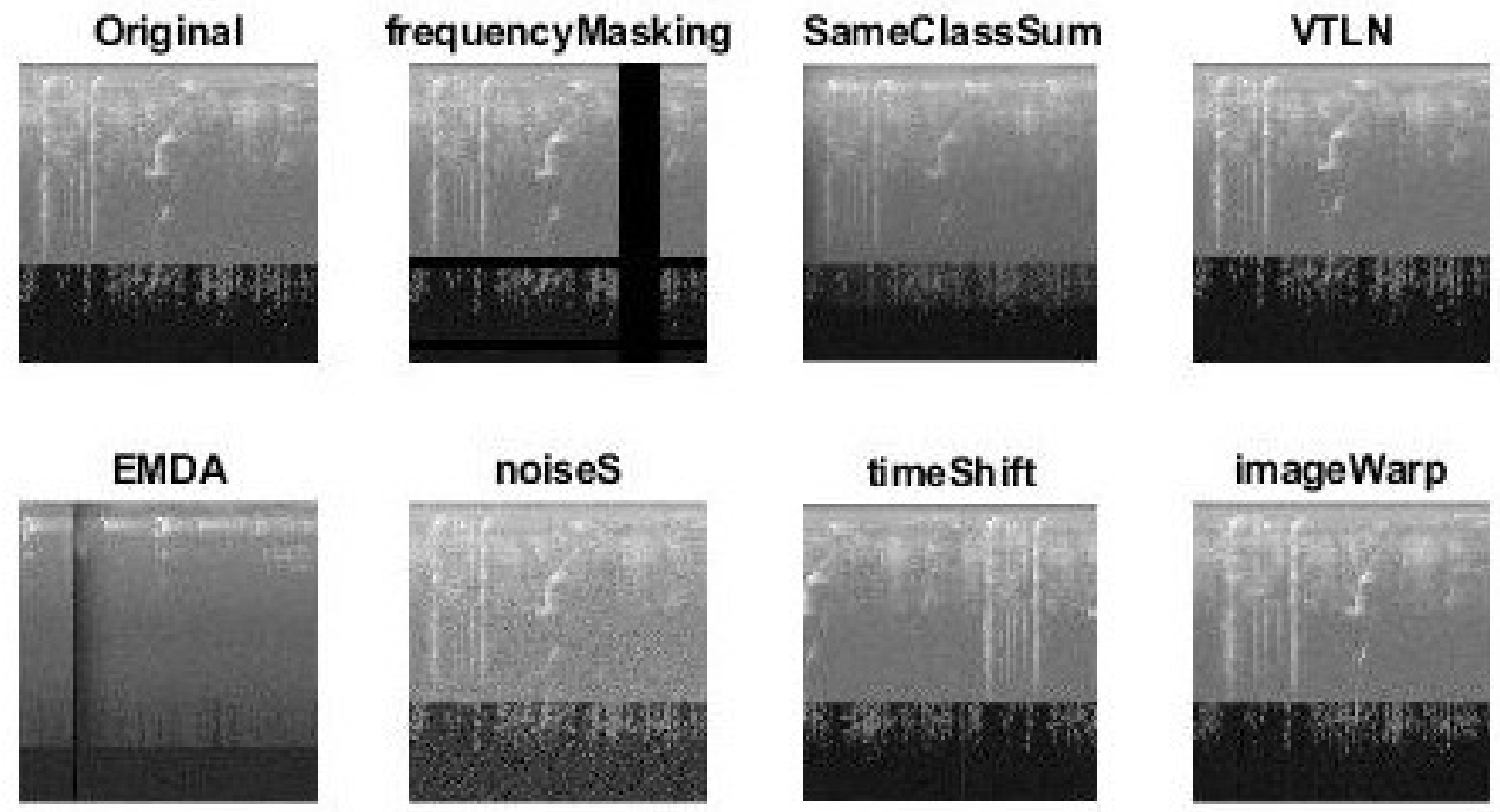

Figure 2. Spectrogram Augmentation. The top left corner shows the spectrogram from the original audio sample. Seven techniques were used to augment the original spectrogram. The description of the outcomes is in the previous section.

In addition, we provide six audio samples from www.xeno-canto.org (original samples in *./examples/OriginalAudioFiles/* and listed as MATLAB table in *./examples/SmallInputDatasets/inputAugmentationFromAudio.mat*) and six spectrograms generated by sgram.m from the aforementioned audio samples (in *./examples/SmallInputDatasets/inputAugmentationFromSpectrograms.mat*). The precomputed results for all the six audio samples and spectrograms are provided in the folder ./examples/AugmentedImages/.

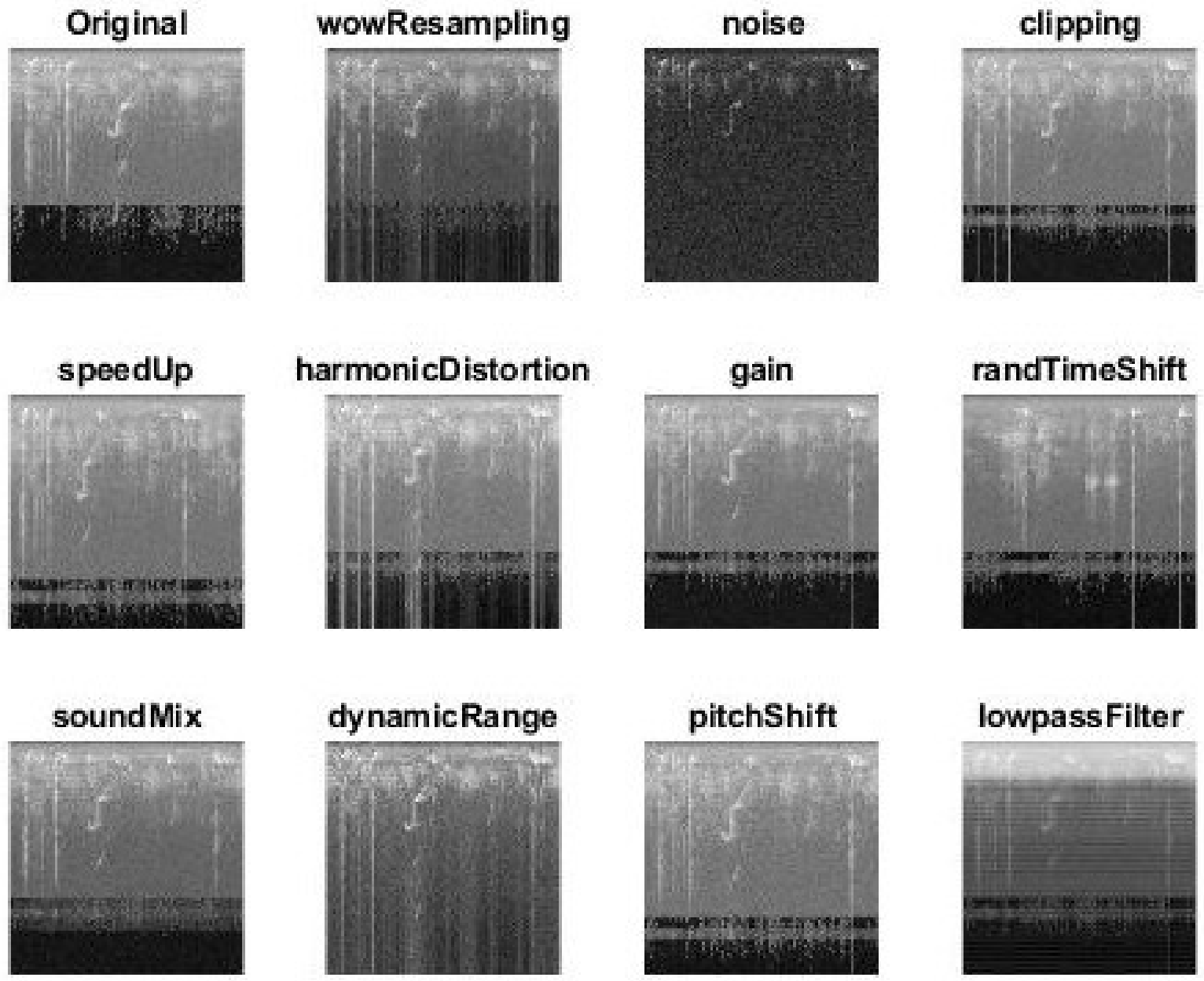

Figure 3. Audio Augmentation. The top left corner shows the spectrogram of the original audio sample. We used 11 audio augmentation methods and extracted the spectrograms. The description of the outcomes is in the previous section.

# 6. Experimental Results

The ESC-50 dataset [27] contain 2000 audio samples evenly divided in 50 classes. These classes are, for example, animal sounds, crying babies and chainsaws. The evaluation protocol proposed by their creators is a five fold cross-validation and the human classification accuracy on this dataset is 81.3%.
We tested seven different augmentation protocol with two different networks, AlexNet [28] and VGG16 [29].

## 6.1 Baseline

The first pipeline is our baseline. We transformed every audio signal into an image representing a Gabor spectrogram. After that we fine-tuned the neural network on the training set of every fold and we evaluated it on their corresponding test set. We trained it with a mini batch of size 64 for 60 epochs. The learning rate was 0.0001, while the learning of last layer was 0.001.

## 6.2 Standard Data Augmentation

The second protocol is the standard MATLAB augmentation. The training works in the same way as the baseline protocol, with the difference that every training set is 10 times larger due to data augmentation. Due to a larger training set, we only used 14 epochs for the training. For every original signal, we created 10 modified signals applying all the following functions:

- Speed up the signal
- Pitch shift application

- Volume gain application
- Random noise addition
- Time shifting

### 6.3 Single Signal Augmentation

The third pipeline consists in applying the audio augmentations to the original signals, and for every augmentation we get a new sample. The training works in the same way as the standard augmentation protocol. We included in the new training set the original samples and 9 modified versions of the same samples obtained by applying

- *applyGain*
- *applyPitchShift*
- *appyRandTimeShift*
- *applySpeedUp*
- *applyWowResampling*
- *applyClipping*
- *applyNoise*
- *applyHarmonicDistortion*
- *applyDynamicRanceCompression*

### 6.4 Single Spectrogram Augmentation

The fourth pipeline consists in applying the audio augmentations to the spectrograms, and for every augmentation we get a new sample. The training works in the same way as the standard augmentation protocol. We included in the new training set the original samples and 5 modified versions of the same samples obtained by applying:

- *applySpectrogramRandomShifts*
- *applyVTLN*
- *applyRandTimeShift*
- *applyRandomImageWarp*
- *applyNoiseS*

### 6.5 Time Scale Modification Augmentation

The fifth pipeline consists in applying the audio augmentations of the Time Scale Modification (TSM) Toolbox to the signals. We refer to the original paper for a description of the algorithms that we use. We apply the following algorithms twice to every signal, once with speed up equal to 0.8, once with that parameter equal to 1.5:

- Overlap Add
- Waveform Similarity Overlap Add
- Phase Vocoder
- Phase Vocoder with Identity Phase Locking

### 6.6 Audio Degradation Toolbox

The sixth augmentation strategy consists in applying 9 techniques that are contained in the Audio Degradation Toolbox (ADT). This works in the same way as Single Signal, but with different algorithms. We applied the following techniques:

- Wow Resampling
- Noise
- Clipping
- Harmonic Distortion
- Sound Mix
- Speed Up
- Aliasing
- Delay
- Lowpass Filter

The results of these protocols are summarized in Table 1.

| | Baseline | Standard | Single Signal (ours) | Single Spectro (ours) | TSM | ADT |
|---|---|---|---|---|---|---|
| AlexNet | 60.80 | 72.75 | 73.85 | 65.75 | 70.95 | 67.65 |
| VGG16 | 71.60 | 79.40 | 80.90 | 75.95 | 79.05 | 77.50 |

Table 1. Classification results of the different protocols

These results show the efficiency of our algorithms, especially when compared to other similar approaches [30,31] that use CNNs with speed up augmentation to classify spectrograms. [30] is a baseline CNN proposed by the creators of the dataset, while in [31] the authors train AlexNet as we do. In both cases, only speed up is used as data augmentation. We outperform both approaches, since they respectively reach 64.5% and 63.2% accuracy. Other networks specifically designed for these problems reach a 86.5%, although using also unlabeled data for training [19]. However, the purpose of these experiments was to prove the validity of the algorithms and the consistency with previous similar approaches. It was not reaching the state of the art performance on ESC-50. We can see that a better performing network like VGG16 nearly reaches human-level classification accuracy, which is 81.3%. The signal augmentation protocol works better than the spectrogram augmentation, but recall that the latter augmentation strategy consists in creating only six new samples. However, Audiogmenter outperforms standard data augmentation techniques when signal augmentation is applied. We do not claim any generalization of the results. The performance of an augmentation strategy depends on the choice of the algorithms, not on its implementation. What we do in our library is proposing a set of tools that must be used smartly by researchers to improve their classifiers performances. We showed in our experiments that Audiogmenter is useful in a very popular and competitive dataset and we encourage researchers to test on different tasks. The code to replicate our experiments can be found in the folder *./demos/.*

## 7. Conclusions

In this paper we proposed Audiogmenter, a novel MATLAB audio data augmentation library. We provide 23 different augmentation methods that work on raw audio signal and their spectrograms. To the best of our knowledge, this is the largest audio data augmentation library in MATLAB. We described the structure of the toolbox and provided examples of its application.

We proved the validity of our algorithm by training a convolutional network on a competitive audio dataset using our data augmentation algorithms and obtained results which are consistent with similar approaches in the literature. The library and its documentation are freely available at https://github.com/LorisNanni/Audiogmenter.